\documentclass[twocolumn,showpacs,preprintnumbers,amsmath,amssymb]{revtex4}
\usepackage{graphicx}
\usepackage{dcolumn}
\usepackage{bm}
\newcommand{\Eq}[1]{Eq.~({\protect\ref{#1}})}

\newcommand{\Ref}[1]{Ref.\protect\cite{#1}}

\newcommand{\Sec}[1]{Sect.~\protect\ref{#1}}
\newcommand{\Fig}[1]{Fig.~\protect\ref{#1}}

\newlength{\Tatescale}
\newlength{\figwidth}
\begin{document}
\title{A statistical approach to the QCD phase transition\\
--- A mystery in the critical temperature}
\author{Noriyoshi Ishii}
\address{
Radiation Laboratory,
The Institute of Physical and Chemical Research (RIKEN),\\
2-1 Hirosawa, Wako, Saitama 351-0198, Japan}

\author{Hideo Suganuma}
\address{Faculty of Science, Tokyo Institute of Technology,\\
2-12-1 Ohokayama, Meguro, Tokyo 152-8551, Japan}


\begin{abstract}
We study the  QCD phase transition based on  the statistical treatment
with the  bag-model picture of hadrons, and  derive a phenomenological
relation among the  low-lying hadron masses, the hadron  sizes and the
critical temperature of the QCD phase transition.
We apply this phenomenological relation  to both full QCD and quenched
QCD,  and compare  these results  with the  corresponding  lattice QCD
results.
%
Whereas such a statistical approach works well in full QCD, it results
in  an extremely  large estimate  of the  critical  temperature in quenched QCD, 
which indicates a serious  problem in understanding of the  QCD phase transition.
This large discrepancy  traces back to the fact  that enough number of
glueballs are  not yet thermally  excited at the  critical temperature
$T_c  \simeq 280$  MeV  in quenched  QCD  due to  the extremely  small
statistical factor as $e^{-m_{\rm{G}}/T_c} \simeq 0.00207$.
This  fact  itself has  a  quite  general  nature independent  of  the
particular  choice of  the  effective model  framework.   We are  thus
arrive  at  a mystery,  namely,  what is  really  the  trigger of  the
deconfinement phase transition.
\end{abstract}
\maketitle
\section{Introduction}
\label{sec.intro}

The quark-gluon-plasma (QGP) is one of the most interesting targets
in the finite-temperature quark-hadron physics \cite{qcd}.  Currently,
the QGP creation experiment is being performed in RHIC project at BNL,
and  much  progress in  understanding  the  finite-temperature QCD  is
desired.
Historically, the instability of the  hadron phase was first argued by
Hagedorn \cite{hagedorn} before the  discovery of QCD. He pointed  out the possibility
of a  phase transition at  finite temperature, based on  the string or the flux-tube
picture of hadrons\cite{hagedorn,patel}.
After  QCD was  established as  the fundamental  theory of  the strong
interaction,    this     transition    was    recognized     as    the
deconfinement phase  transition to the QGP  phase, where quarks
and gluons are liberated with the restored chiral symmetry.
The  QCD phase  transition  has been  studied  using various  infrared
effective   models    of   QCD    such   as   the    linear   $\sigma$
model\cite{kapusta},               the              Nambu-Jona-Lasinio
model\cite{hatsuda-kunihiro},     the    dual    Ginzburg-Landau
theory\cite{ichie} and so on.

In  order   to  study  nonperturbative  features  of   the  QCD  phase
transition,  the  lattice QCD  Monte  Carlo  calculation  serves as  a
powerful tool directly based on QCD.
It has  been already extensively used  to study the nature  of the QCD
phase transition.  At the  quenched level, SU(3) lattice QCD indicates
the existence  of the deconfinement  phase transition of a  weak first
order at $T_c \simeq 260-280$  MeV \cite{karsch2}.  On the other hand,
in the  presence of  dynamical quarks, it  indicates the  chiral phase
transition at  $T_c = 173(3)$ MeV  for $N_f=2$ and $T_c  = 154(8)$ MeV
for $N_f = 3$ in the chiral limit \cite{karsch}.

In this paper,  we attempt to understand the  physical implications of
the recent lattice QCD results on the QCD phase transition,
and point out an abnormal  nature of the quenched QCD phase transition
based on the statistical argument.
In the  actual calculation, we  adopt the statistical approach  to the
QCD   phase  transition   with  the   bag-model  picture   of  hadrons
\cite{jaffe}.   Although   it  is  a   simply-minded  phenomenological
effective  model, it  can reproduce  the critical  temperature  of the
full-QCD phase transition ``amazingly'' well.
%
In  spite of this  success, this  approach terribly  overestimates the
critical temperature in quenched QCD.
We consider the essential cause  of this overestimate, and point out a
serious   problem  hidden   in   the  QCD   phase   transition  in   a
model-independent manner.

The contents are organized  as follows. In \Sec{sec.full.qcd}, we give
a brief review of the statistical approach to the QCD phase transition
with the  bag-model picture of hadrons, and  derive a phenomenological
relation among  the hadron masses,  the hadron sizes and  the critical
temperature.  We apply this relation to full QCD.
In  \Sec{sec.quench.qcd},  we  apply  this relation  to  quenched  QCD
without  dynamical quarks,  and point  out  a serious  problem on  the
critical temperature.   In \Sec{sec.discussion}, after  the summary of
the results,  we attempt to clarify  the essence of  this problem, and
discuss an  abnormal nature of  the deconfinement phase  transition in
quenched QCD.

\section{A statistical approach to the full-QCD phase transition}
\label{sec.full.qcd}

We investigate  analytically the features of the  QCD phase transition
based  on the  statistical  treatment with  the  bag-model picture  of
hadrons.  We begin by deriving a phenomenological relation between the
critical temperature  $T_c$ and the  properties of the  hadrons, i.e.,
the mass  and the size.  In  the bag-model picture,  quarks and gluons
are  assumed to  be  confined  inside a  spherical  bag.  Here,  color
confinement  is  simply  taken   into  account  through  the  bag-like
intrinsic structure of hadrons.
At  low temperature,  only a  small number  of such  bags  are thermally
excited, and the thermodynamic properties of the system are described in
terms of these spatially isolated bags.
 With the increasing temperature,  the number of the thermally excited
bags increases.   Gradually, these bags  begin to overlap  one another,
and  they  finally  cover  the   whole  space  region  at  a  critical
temperature $T_c$ in this picture.
Above $T_c$,  since the  volume of the  outside disappears,  the whole
space is  filled with liberated quarks and gluons, and the thermodynamic
properties of the system is now governed by quarks and gluons.  
In this  way, the QCD phase transition  is described in terms
of the  overlaps of the thermally  excited bags.
Such a physical interpretation of the QCD phase transition 
is called as the closed packing picture.

\begin{figure}
\includegraphics[width=\figwidth]{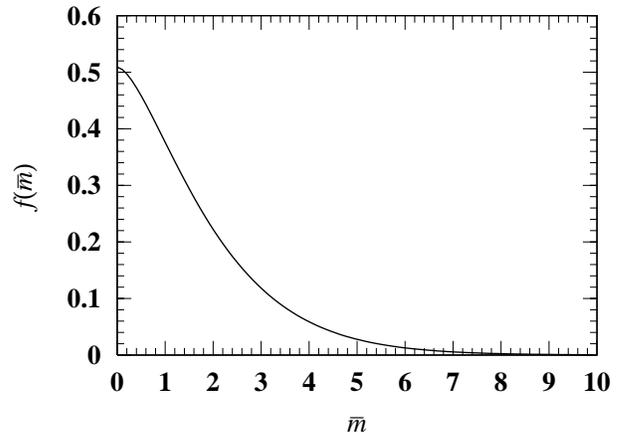}
\caption{The function  $f(\bar m)$  in \Eq{fmt} plotted  against $\bar
m=m/T$.}
\label{f-m-t}
\end{figure}

To  proceed,  we  define  the  spatial occupation  ratio  $r_V(T)$  at
temperature $T$  to be the  ratio of the  total volume of  the spatial
regions inside  the thermally  excited bags to  the volume $V$  of the
whole space  region.  In the closed  packing picture of  the QCD phase
transition, $r_V(T)$ plays the key role, which is estimated as
\begin{eqnarray}
	r_V(T)
&=&
	\frac1V \sum_n{4\pi\over  3} R_n^3 \cdot  \lambda_n N_n(T) 
\label{rv}
\\\nonumber
&=&
	\sum_n 
	\lambda_n 
	R_n^3
	{4\pi \over 3}
	\int {d^3k\over(2\pi)^3}
	{1\over e^{\sqrt{m_n^2 + \vec k^2}/T} - 1} 
\\\nonumber
&=&	\sum_n 
	\lambda_n 
	R_n^3 T^3 f(m_n/T),
\end{eqnarray}
where  $N_n(T)$,  $\lambda_n$,  $m_n$  and  $R_n$ are  the  number  at
temperature $T$,  the degeneracy, the mass  and the bag  radius of the
$n$-th elementary  excitation, respectively. 
Here, $f(\bar m)$ is defined by
\begin{eqnarray}
	f(\bar m)
&\equiv&
	{4\pi \over 3}
	\int {d^3\bar k\over(2\pi)^3}
	{1\over e^{\sqrt{\bar m^2 + \bar k^2}} - 1},
\label{fmt}
\end{eqnarray}
and  its functional form  is plotted  against $\bar  m \equiv  m/T$ in
\Fig{f-m-t}.   Note  that $f(m/T)$  is  a  characteristic function  to
describe the  thermal contribution of the  boson with the  mass $m$ at
the temperature  $T$ \cite{ST91}.  For  $m \gg T$,  $f(m/T)$ decreases
exponentially with $m/T$, and  the thermal contribution is expected to
become negligible.

As mentioned before, in the closed packing picture, the phase transition
takes place,  when the thermally  excited bags almost cover  the whole
space region.   Hence, the critical temperature $T_c$  is estimated by
solving
\begin{equation}
	r_V(T_c) = 1.
\label{critical.temperature}
\end{equation}



Now,  we  investigate the  phase  transition  in  full QCD  using  the
statistical approach.   In full QCD, the  lightest physical excitation
is the  pion, and  all the other  hadrons are  rather heavy as  $m \gg
m_\pi$, $T_c$.  In  fact, the pion is considered to  play the key role
in  describing the thermodynamic  properties of  full QCD  below $T_c$
from the viewpoint of the statistical physics.
%
Hence, in  most cases,  only the pionic  degrees of freedom  are taken
into account in the hadron phase in the argument of the full-QCD phase
transition.
%
By using the isospin degeneracy $\lambda_{\pi} = 3$, the mass $m_{\pi}
=  140$  MeV   and  the  radius  $R_{\pi}  \simeq   1$  fm,  we  solve
\Eq{critical.temperature}  with $r_V(T)=3 R_{\pi}^3  T^3 f(m_{\pi}/T)$
to estimate the critical temperature as $T_c \simeq 183$ MeV.
Considering its closeness to the full lattice QCD result with $N_f=2$,
i.e., $T_c \simeq  170$ MeV, the statistical approach  to the full-QCD
phase transition seems to be rather good.

We now  consider the $m_{\pi}$-dependence of  the critical temperature
$T_c$. Note  that, in the  actual lattice QCD calculations,  the pion
mass  is  taken  to  be  still  rather heavy  as  $m_{\pi}  \agt  400$
MeV for the technical reasons. 
From these data, 
the critical temperature $T_c$ in the chiral limit is obtained 
using the chiral extrapolation.
\begin{figure}
\includegraphics[width=\figwidth]{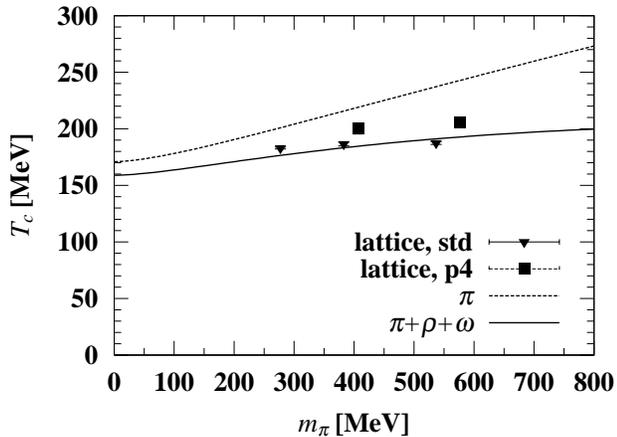}
\caption{ The critical temperature $T_c$ plotted against the pion mass
$m_{\pi}$ in  the $N_f=2$ case.   The dashed curve denotes  the result
retaining only  the contribution from pions.  The  solid curve denotes
the  result  including also  contributions  from  $\rho$ and  $\omega$
mesons.    The  triangle   denotes   the  lattice   data  taken   from
\Ref{luetgemeier}  obtained with standard  (std) fermion  action.  The
square denotes the lattice  data taken from \Ref{karsch} obtained with
the improved staggered (p4) fermion action. }
\label{tc.pion}
\end{figure}
In \Ref{karsch}, the authors  parametrized the full lattice QCD result
of $T_c$ as
\begin{equation}
	\left( {T_c \over \sqrt{\sigma}}\right)(m_{\pi})
=
	0.40(1)
+
	0.039(4)\left( {m_{\pi} \over \sqrt{\sigma}} \right),
\label{parameterize}
\end{equation}
where  $\sigma$ denotes  the string  tension.  
Strictly speaking, the phase  transition becomes just a cross-over for
intermediate values  of $m_{\pi}$. Hence, in  \Ref{karsch}, the pseudo
critical temperature is adopted as $T_c$, which is determined from the
peak positions of the susceptibilities of the Polyakov loop and so on.

For  the $N_f=2$  case,  we plot,  in  \Fig{tc.pion}, the  theoretical
estimate  of  $T_c$  against   $m_{\pi}$  based  on  \Eq{fmt}  in  the
statistical  approach with  the bag-model  picture.  The  dashed curve
denotes $T_c$  retaining only the contribution from  pions with $R_\pi
\simeq 1$ fm.
%
%
%
The  solid curve denotes  the result  in the  case including  also the
contributions  from the low-lying  vector mesons,  such as  $\rho$ and
$\omega$, which are  the next lightest particles in  $N_f=2$ full QCD.
Here, we have used $\lambda_{\rho}=3\times 3=9$, $m_{\rho} = 770$ MeV,
$R_{\rho}\simeq  1$ fm,  $\lambda_{\omega}= 3$,  $m_{\omega}=783$ MeV,
$R_{\omega}\simeq   1$   fm   as   inputs,  which   are   treated   as
$m_{\pi}$-independent   constants.   These   vector  mesons   give  an
additional contribution  to $r_V(T)$ as $\delta r_V(T)  = 9 R_{\rho}^3
T^3 f(m_{\rho}/T) + 3 R_{\omega}^3 T^3 f(m_{\omega}/T)$.
The triangle and  the square in \Fig{tc.pion} denote  the lattice data
taken from Refs.\cite{luetgemeier,karsch}.
%
In spite of its roughness,
the  resulting behavior  of the  critical temperature  $T_c$  seems to
coincide  rather   well  with  the   lattice  QCD  result.    In  this
calculation, the low-lying vector  mesons, $\rho$ and $\omega$, give a
small but non-negligible contribution to the critical temperature.

\begin{figure}
\includegraphics[width=\figwidth]{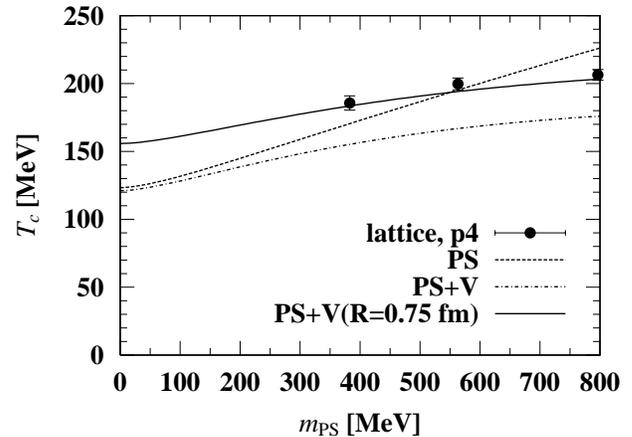}
\caption{The   critical   temperature   $T_c$  plotted   against   the
pseudo-scalar  meson  mass $m_{\rm{PS}}$  in  the SU(3)$_f$  symmetric
case.   The  dashed  curve  denotes  the  result  retaining  only  the
contribution of the pseudo-scalar mesons. The dot-dashed and the solid
curves  denote the  results  including also  the  contribution of  the
flavor-octet  vector  mesons  with  $R   =  1$  fm  and  $R=0.75$  fm,
respectively.   The  circle  denotes   the  lattice  data  taken  from
\Ref{karsch}  obtained  with   the  improved  staggered  (p4)  fermion
action.}
\label{tc.pseudo-scalar}
\end{figure}
We next investigate the  idealized SU(3)$_f$ symmetric case to analyze
the  other lattice  QCD data.   In this  case, the  pseudo-scalar (PS)
octet mesons, such as pions,  kaons and $\eta_8$, are the lightest and
possess the  same mass $m_{\rm PS}$  in common with  the degeneracy as
$\lambda_{\rm{PS}}=8$.    We  plot,  in   \Fig{tc.pseudo-scalar},  the
critical  temperature $T_c$ against  $m_{\rm{PS}}$.  The  dashed curve
denotes  $T_c$ retaining  only  the contribution  from PS-mesons  with
$R_{\rm PS} \simeq 1$ fm, which leads to $r_V(T) = 8 R_{\rm{PS}}^3 T^3
f(m_{\rm{PS}}/T)$.   The  dot-dashed  curve in  \Fig{tc.pseudo-scalar}
denotes the  results including also  the contributions from  the octet
vector    mesons    with    $\lambda_{\rm{V}}   =    8\times    3=24$,
$m_{\rm{V}}\simeq 770$ MeV, $R_{\rm{V}}\simeq 1$ fm.
%
(Inclusion  of the  flavor-singlet vector  meson does  not  change the
result so much.)
These  vector mesons give  an additional  contribution to  $r_V(T)$ as
$\delta r_V(T)  = 24  R_{\rm{V}}^3 T^3 f(m_{\rm  V}/T)$.
%
The circle  in \Fig{tc.pseudo-scalar}  denotes the lattice  data taken
from \Ref{karsch}.
We  see that both  the dashed  and the  dot-dashed curves  are roughly
consistent with the lattice QCD results.
%
We  note  that  the  small  deviation almost  disappears  by  slightly
adjusting the bag  size as $R_{\rm{PS}} = R_{\rm{V}}  =0.75$ fm, as is
shown in \Fig{tc.pseudo-scalar} with the solid curve.

In  this way,  this  simple statistical  approach  with the  bag-model
picture  works rather  well  in reproducing  the critical  temperature
$T_c$ of  the full-QCD phase transition.  Here,  the main contribution
is given by  the thermal pion, and the low-lying  vector mesons give a
small but non-negligible contribution to $T_c$.

\section{A Mystery in the Quenched QCD Phase Transition}
\label{sec.quench.qcd}

In spite of the absence of the dynamical quarks, quenched QCD provides
various   important  nonperturbative  features   such  as   the  color
confinement  and  instantons.   Hence,   to  understand  some  of  the
nonperturbative natures  of QCD, quenched QCD plays  the primary role,
serving as a simplified version  of the complicated real problems.  In
this  section, we apply  the statistical  approach with  the bag-model
picture  to quenched  QCD in  a similar  manner done  in  the previous
section.


Due to the color confinement,  only the color-singlet modes can appear 
as  physical  excitations, and such modes are called as glueballs in quenched QCD.
The mass  spectrum of the glueballs is known through the quenched lattice QCD calculations
\cite{morningstar,weingarten,teper}.  The lightest physical excitation 
is the  $0^{++}$ glueball with  $m_{\rm G} = 1.5-1.7$  GeV.  
Following the similar argument as  in the  full QCD  case, the  lowest $0^{++}$ 
glueball is expected to play the key  role  in  describing the  thermodynamic 
properties of quenched  QCD in the confinement phase.   Hence, we first take
into account only the $0^{++}$ glueball.
As for the size $R$ of the scalar glueball, we note that there is no widely agreed value on it. 
We adopt a rather small value as $R \sim$ 0.4fm, which has been indicated by recent lattice QCD studies \cite{ishii.full.paper,ishii}.

We use the degeneracy $\lambda_{\rm G(S)} = 1$, the mass $m_{\rm G(S)}
\simeq 1730$ MeV \cite{morningstar},  and $R_{\rm G(S)} \simeq 0.4$ fm
as inputs.  Then,  the spatial occupation ratio is  given by $r_V(T) =
R_{\rm{G(S)}}^3 T^3f(m_{\rm G(S)}/T)$, and the critical temperature is
estimated  as  $T_c \simeq  827$  MeV from  \Eq{critical.temperature}.
This estimate is too much  larger than the quenched lattice QCD result
as $T_c = 0.629(3)\sqrt{\sigma}  \simeq 280$ MeV in \Ref{karsch2} with
$\sqrt{\sigma} = 450$  MeV.  (In other words, only  a tiny fraction of
the space region is covered by the thermally excited bags as $r_V(T) =
0.0021$ at  $T=280$ MeV.)  This  large discrepancy would be  a serious
problem in the quenched QCD phase transition.

To seek a  possible solution, we examine the case  with a larger value
of the glueball size $R$, since there is no established value on $R$.
\begin{figure}
\includegraphics[width=\figwidth]{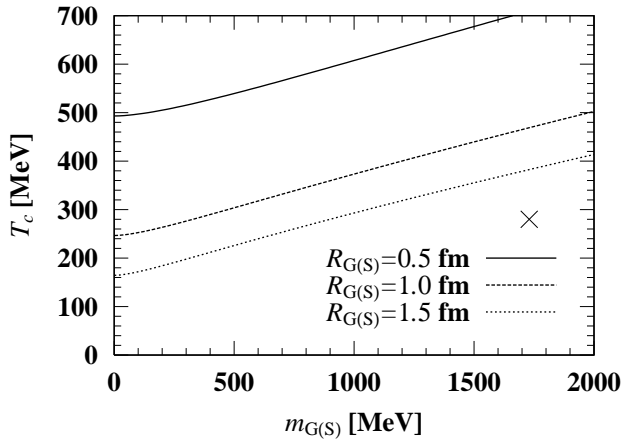}
\caption{ The  critical temperature  $T_c$ plotted against  the lowest
$0^{++}$  glueball mass  $m_{\rm{G(S)}}$ in  the  statistical approach
assuming various  glueball sizes as $R_{\rm{G(S)}}=0.5,  1.0, 1.5$ fm.
The  cross  ($\times$) indicates  the  quenched  lattice QCD  results,
$T_c=280$ MeV and $m_{\rm G(S)}=1730$ MeV.  }
\label{tc-mG}
\end{figure}
In \Fig{tc-mG}, the critical  temperature $T_c$ is plotted against the
lightest $0^{++}$  glueball mass $m_{\rm{G(S)}}$ in the statistical approach 
for various glueball sizes as $R_{\rm{G(S)}} = 0.5, 1.0, 1.5$ fm.
The  cross  ($\times$) indicates  the  quenched  lattice QCD  results,
$T_c=280$ MeV and $m_{\rm G}=1730$ MeV.
In this argument, to reproduce the critical temperature as $T_c \simeq
280$  MeV,   the  glueball   size  is  to   be  abnormally   large  as
$R_{\rm{G(S)}} \simeq 3.1$ fm in the vicinity of $T_c$.  However, such
a drastic thermal swelling of  the lowest scalar glueball was rejected
by  the  recent lattice  QCD  studies \cite{ishii.full.paper,  ishii},
which states that  the thermal glueball size is  almost unchanged even
near $T_c$.
In any case, the glueball size does not seem to provide the solution on this discrepancy.

We may seek for the solution in the drastic pole-mass reduction of the
$0^{++}$ glueball  near the critical  temperature as was  suggested in
\Ref{ichie}  in the  context of  the dual  Ginzburg-Landau theory.  In
\Fig{tc-mG}, we see that, for the problem to be settled, the pole-mass
reduction must be as significant as $m_{\rm{G(S)}}(T_c) \alt 500$ MeV.
However,     in     the     recent    lattice     QCD     calculations
\cite{ishii.full.paper,ishii}, it  has been reported  that the thermal
$0^{++}$ glueball persists to hold a rather large pole-mass as $m_{\rm
G(S)}(T  \simeq T_c) \simeq  1250$ MeV.   Hence, we  have to  seek for
another possibility by including the contribution of the excited-state
glueballs.

In addition  to the lowest $0^{++}$ glueball  with $m_{\rm G(S)}\simeq 1730$
MeV and  $R_{\rm G(S)} \simeq 0.4$ fm,  we consider the  thermal contribution
from the lowest  $2^{++}$ glueball, which is the  next lightest hadron
in quenched QCD.
%
We  take  $\lambda_{\rm{G(T)}}=5$,  $m_{\rm{G(T)}}  \simeq  2400$  MeV
\cite{morningstar}, $R_{\rm{G(T)}}\simeq 1$ fm. (Here, we assume it to
have a typical hadron size.)
Then, the  spatial occupation ratio  receives a correction  as $\delta
r_V(T) = 5  R_{\rm{G(T)}}^3 T^3f(m_{\rm{G(T)}}/T)$, and the correction
amounts to  $\delta r_V(T)  = 0.0223$ at  $T=280$ MeV.   The resulting
critical temperature is given as  $T_c \simeq 432$ MeV, which is still
too large.
We   note   that  the   realistic   glueball   size   would  be   more
compact. However, if so, its contribution becomes more negligible.

Besides these  two low-lying  glueballs, the following  excited states
are predicted in \Ref{morningstar} as $0^{-+}(2590)$, $0^{*++}(2670)$,
$1^{+-}(2940)$,   $2^{-+}(3100)$,   $3^{+-}(3550)$,   $0^{*-+}(3640)$,
$3^{++}(3690)$,   $1^{--}(3850)$,   $2^{*-+}(3890)$,   $2^{--}(3930)$,
$3^{--}(4130)$, $2^{+-}(4140)$, $0^{+-}(4740)$.
%
We include  all the contribution  from these excited  states, assuming
the  unknown glueball  size as  a typical  hadron size,  i.e., $R_{\rm
G}\simeq 1$  fm.  The  correction by these  excited states  amounts to
only  $\delta r_V(T)  = 0.0113$  at $T=  280$ MeV,  and  the resulting
critical temperature is  estimated as $T_c = 395$  MeV, which is again
too large.
%

In this way, we observe that the statistical approach does not work at
all  in  quenched  QCD.  The  direct  cause  of  this failure  is  the
extremely  small  statistical  factor  $e^{-m_{\rm{G(S)}}/T_c}  \simeq
0.0021$, which  strongly suppresses the excitations  of the glueballs.
As a consequence, only a  remarkably tiny fraction of the space region
can be covered  by the thermally excited bags  of glueballs at $T=280$
MeV, which cannot be the driving force of the QCD phase transition.

\section{Summary and Discussions 
---What is the trigger or the driving force of the QCD phase transition ?}
\label{sec.discussion}

We have  analytically studied  the QCD phase  transition based  on the
statistical treatment  with the bag-model picture.  We  have derived a
phenomenological  relation among the  critical temperature  $T_c$, the
mass and  the size of the  low-lying hadrons.  First,  we have applied
this relation  to full QCD,  and have compared the  analytical results
with the  full lattice QCD results.   In full QCD, we  have found that
this  approach  works amazingly  well.   As  is  expected, the  pionic
contribution seems to  play the main role in  determining the critical
temperature, and  the low-lying vector  mesons are found to  provide a
small but non-negligible contribution.

Unlike full QCD, we have  found that the statistical approach terribly
overestimates  the critical  temperature as  $T_c \simeq  827$  MeV in
quenched QCD,  and that only  a tiny fraction  of the space  region is
covered  by  the thermally  excited  bags  of  glueballs at  $T\simeq$
280MeV.
%
%
We have considered the possibility  of a swelling of the glueball size
and  reduction of  the glueball  mass near  the  critical temperature.
However, both of  these two have not provided us  with the solution on
the large discrepancy.
%
In  spite of  including all  the contributions  from the  15 low-lying
glueballs   up  to   5   GeV  predicted   in   quenched  lattice   QCD
\cite{morningstar},  the discrepancy  remains to  be still  large.  In
other words,  the number of  the thermally-excited glueballs  is still
too small  at $T \simeq$ 280MeV,  even after so  many glueball excited
states are taken into account.
The direct origin of this discrepancy is the strong suppression of the
thermal excitation of glueballs due to the extremely small statistical
factor as  $e^{-m_{\rm{G(S)}}/T} = 0.00207$  at $T=280$ MeV even  for the
lightest  glueball, which  leads  to the  insufficient  amount of  the
covered space by  the thermally excited bags of  glueballs.

Although  several  arguments given  so  far  have  been based  on  the
bag-model  picture, this problem  itself has  a quite  general nature,
which can go beyond the reliability of the model framework.
We  finally reformulate  this problem  in a  model-independent general
manner by  considering the inter-particle distance $l$  instead of the
bag size $R$.
%
Note that the reliability of the statistical argument becomes improved
in the  dilute glueball gas limit.   Now, taking into  account all the
low-lying 15 glueball modes up  to 5 GeV predicted in quenched lattice
QCD \cite{morningstar},  we calculate the  inter-particle distance $l$
of  the  glueballs based on only the statistical argument.
At $T=280$  MeV, the inter-particle distance is  estimated as $l\simeq
5$ fm.
It  follows that  the deconfinement  phase transition  takes  place at
$T\simeq$  280MeV,   when  the   glueball  density  becomes   $\rho  =
1/\{\frac{4\pi}{3}(2.5{\rm  fm})^3\} \simeq 1/(4.0{\rm  fm})^3$.
Remember  that the  theoretical  estimates of  the  glueball size  are
rather      small     as      $R_{\rm{G(S)}}     \alt      0.4$     fm
\cite{ishii,ishii.full.paper,schafer}.   Hence, this density  seems to
be rather dilute.
%
Since the  long-range interaction among  glueballs is mediated  by the
virtual   one-glueball   exchange  process   in   quenched  QCD,   the
interactions among  glueballs are exponentially  suppressed beyond its
Compton  length $1/m_{\rm{G(S)}}  = 0.112$  fm. Therefore,  one cannot
expect   the   strong   long-range   interaction  acting   among   the
spatially-separated thermal glueballs.

In this  way, it  is quite difficult  to imagine  how such a  too rare
excitation of thermal  glueballs can lead to the  phase transition. We
are thus  arrive at  the mystery.  What is really  the trigger  or the
driving force of the deconfinement  phase transition in quenched QCD ?
In order to  understand the QCD phase transition,  this problem should
be seriously considered.




%
%

\hspace{0.5em}

\begin{center}{\bf Acknowledgement}\end{center}
H.~S. is supported by Grant for Scientific Research (No.12640274) from
Ministry    of   Education,    Culture,   Science    and   Technology,
Japan.


\begin{references}
\bibitem{qcd}
			W. Greiner and A. Sch\"afer,
			``Quantum Chromodynamics'',
			(Springer-Verlag, Berlin, 1994) 1.
\bibitem{hagedorn}
			R.~Hagedorn, N.~Cim. {\bf 56A}, 1027 (1968). 
\bibitem{patel}
			A.~Patel, Nucl.~Phys. {\bf B243}, 411 (1984);
			Phys.~Lett. {\bf B139}, 394 (1984).
\bibitem{kapusta}
			J.~Kapusta,
			``Finite-Temperature Field Theory'',
			Cambridge Monographs on 
			Mathematical Physics, (1989) 1.
\bibitem{hatsuda-kunihiro}
			T.~Hatsuda and T.~Kunihiro,
			Phys. Rep. {\bf 247}, 221 (1994).
\bibitem{ichie}
			H. Ichie, H. Suganuma and H. Toki,
			Phys. Rev. {\bf{D52}}, 2944 (1995).
\bibitem{karsch2}
			G. Boyd, J. Engels, F. Karsch,
			E. Laermann, C. Legeland,
			M. Lutgemeier, B. Petersson,
			Nucl. Phys. {\bf{B469}}, 419 (1996).
\bibitem{karsch}
			F. Karsch, E. Laermann and A. Peikert,
			Nucl. Phys. {\bf B605}, 579 (2001).
\bibitem{jaffe}
			A.~Chodos, R.~L.~Jaffe, K.~Johnson and C.~B.~Thorn,
			Phys. Rev. {\bf D10}, 2599 (1974).
%
\bibitem{ST91}
			H.~Suganuma and T.~Tatsumi,
			Ann. Phys. {\bf 208}, 470 (1991).
%
\bibitem{luetgemeier}
			M.~L\"{u}tgemeier, Phasen\"{u}berg\"{a}nge in der
			QCD mit fundamentalen und adjungierten Quarks,
			Ph.D. thesis, Bielefeld, (1998).
\bibitem{morningstar}
			C. J. Morningstar and M. Peardon,
			Phys. Rev. {\bf D60}, 034509 (1999),
			and references therein.
\bibitem{weingarten}
			J. Sexton, A.Vaccarino and D. Weingarten,
			Phys. Rev. Lett. {\bf 75}, 4563 (1995),
			and references therein.
\bibitem{teper}
			M. J. Teper,
			OUTP-98-88-P (1998), hep-th/9812187.
\bibitem{ishii}
			N. Ishii, H. Suganuma and H. Matsufuru,
			Phys. Rev. {\bf{D66}}, 014507 (2002).
%
\bibitem{ishii.full.paper}
			N. Ishii, H. Suganuma and H. Matsufuru,
			hep-lat/0206020, Phys. Rev. {\bf D66}, 0745XX (2002) in press.
%
%
%
\bibitem{schafer}
			T.~Sch\"{a}fer and E.~Shuryak,
			Phys. Rev. Lett. {\bf 75}, 1707 (1995).
\end{references}
\end{document}